\documentclass[journal]{IEEEtran}
\usepackage{amsmath,amssymb}
\usepackage{graphicx}
\usepackage{epstopdf}
\usepackage{color}
\usepackage{subcaption}
\usepackage{stfloats}
\usepackage{caption}
\usepackage{array}
\usepackage{enumerate}
\usepackage{makecell}
\usepackage{colortbl}
\usepackage{hyperref}
\usepackage{cleveref}
\usepackage[stable]{footmisc}

\usepackage{fancyhdr}
\usepackage{tikz}
\newcommand\copyrighttext{%
  \footnotesize \textcopyright 2017 IEEE. Personal use of this material is permitted.
  Permission from IEEE must be obtained for all other uses, in any current or future
  media, including reprinting/republishing this material for advertising or promotional
  purposes, creating new collective works, for resale or redistribution to servers or
  lists, or reuse of any copyrighted component of this work in other works.
  DOI: \href{https://doi.org/10.1109/ISCAS.2017.8050240}{10.1109/ISCAS.2017.8050240}}
\newcommand\copyrightnotice{%
\begin{tikzpicture}[remember picture,overlay]
\node[anchor=south,yshift=0pt] at (current page.south) {\fbox{\parbox{\dimexpr\textwidth-\fboxsep-\fboxrule\relax}{\copyrighttext}}};
\end{tikzpicture}%
}

\usepackage[mathlines,switch]{lineno}

\begin{document}
		\title{
			Non-invasive Blood Pressure Estimation Using Phonocardiogram
		}

\author{
\thanks{\textbf{The collected data set can be accessed using the following url link: \url{www.kaggle.com/mkachuee/noninvasivebp}.}}

\IEEEauthorblockN{Amirhossein Esmaili, Mohammad Kachuee, Mahdi Shabany\\}
	\IEEEauthorblockA{Department of Electrical Engineering\\
		Sharif University of Technology\\
		Tehran, Iran\\
		Email: amirhossein.ed12@gmail.com, m.kachuee@gmail.com, mahdi@sharif.edu
		}
		}

\maketitle
\copyrightnotice

\begin{abstract}
This paper presents a novel approach based on pulse transit time (PTT) for the estimation of blood pressure (BP). \textcolor[rgb]{0.00,0.00,0.00}{In order to achieve this goal, a data acquisition hardware is designed for high-resolution sampling of phonocardiogram (PCG) and photoplethysmogram (PPG). These two signals can derive PTT values.} Meanwhile, a force-sensing resistor (FSR) is placed under the cuff of the BP reference device to mark the moments of measurements accurately via recording instantaneous cuff pressure. For deriving the PTT-BP models, a calibration procedure including a supervised physical exercise is conducted for each individual. The proposed method is evaluated on 24 subjects. 
The final results prove that using PCG for PTT measurement alongside the proposed models, the BP can be estimated reliably. Since the use of PCG requires a minimal low-cost hardware, the proposed method enables ubiquitous BP estimation in portable healthcare devices.
\end{abstract}

\begin{IEEEkeywords}
cuff-less blood pressure, phonocardiogram (PCG), pulse transit time (PTT), pulse arrival time (PAT), mobile health (mHealth)
\end{IEEEkeywords}

\section{Introduction} \label{section.introduction}
Hypertension is one of the main reasons that increases the risk of cardiovascular diseases.
Prevalence of raised blood pressure (BP) among adults is 26.4 percent for men and 23.1 percent for women worldwide and it is more pronounced in low-income countries \cite{world2015world}.
BP is a periodic signal
with the heart rate frequency and defined as the pressure applied on the vessel walls
when the blood circulates throughout the body. The upper limit of BP is called
systolic blood pressure (SBP) and its lower limit is called diastolic blood pressure (DBP). 

Currently, auscultatory and oscillometry are the most common approaches for non-invasively BP measurement. 
However, these two methods require the use of a inflating cuff around the arm, which causes inconvenience for patients and also does not allow continuous BP monitoring. Therefore, since the patient's BP may change over time due to various mental or physical factors, a cuff-less non-invasive method, which can monitor BP continuously, would be desirable.

Various research efforts have been performed trying to provide
methods for cuff-less and continuous BP measurement, among which using cardiovascular
surrogate parameters, especially the pulse wave velocity (PWV),
is the most common approach. In fact, PWV is the velocity of a
pressure wave propagation in vessels and can be estimated
by the time that it takes for the pressure wave to travel from
one point of the arterial tree (proximal point) to another one (distal point) within the same cardiac cycle. This time interval is called pulse transit time (PTT).




The main principle behind using PWV for BP estimation is that the blood flow in the arteries can be modeled as the propagation of pressure waves inside elastic tubes.
According to \cite{kachuee2016cuff}, by assuming vessels like elastic tubes, the following formulation for PTT can be derived:
\begin{equation}\label{eq:PWV_2}
PTT = l \sqrt{ \frac{\rho A_{m}}{\pi A P_{1} [1+(\frac{P-P_{0}}{P_{1}})^{2}]} } ,
\end{equation}
where $P$ is the fluid pressure (here BP), $\rho$ is the blood density, $A$ is the vessel cross section and $P_0$, $P_1$ and $A_m$ are individual-specific physical parameters. According to \eqref{eq:PWV_2}, the exact relation bewteen BP and PTT depends on the physical properties of the vessels and blood. Therefore, a calibration procedure is required for each individual to approximate this relation. 

By solving (\ref{eq:PWV_2}) for $P$, BP can be written as a function of PTT:
\begin{equation}\label{eq:P}
BP = a_{0} + \sqrt{a_{1} + a_{2} \frac{1}{PTT^{2}}},
\end{equation}
where $a_{0}$, $a_{1}$, and $a_{2}$ are constant coefficients, which are functions of the individual-specific parameters of (\ref{eq:PWV_2}). 
\textcolor[rgb]{0.00,0.00,0.00}{(\ref{eq:P}) can be approximated by the following formulation:}
\begin{align}
&BP=b_0 + b_1PTT^{-1}, \label{eq.pwv_inv_bp}
\end{align}
which describes a simple model for estimating BP. \textcolor[rgb]{0.00,0.00,0.00}{In this paper, the PTT-BP relation in (\ref{eq.pwv_inv_bp}) is used for the estimation of individual's BP. In order to fit this model, a linear regression with mean squared error cost function is used to calculate the subject-specific parameters of the model.}


Futhermore, the majority of research papers in literature, exploit the electrocardiogram (ECG) and the photoplethysmogram (PPG) for the PTT calculation \cite{gesche2012continuous}, 
\cite{cattivelli2009noninvasive}.
In fact, using these two signals provides another
cardiovascular parameter, which is called pulse arrival time (PAT)
instead of PTT, where PAT includes pre-ejection period (PEP) in
addition to PTT. Although PTT is related to BP, PEP can be affected by some factors
other than variations in BP like myocardial
contractility \cite{talley1971evaluation}. Furthermore,
acquiring ECG signals requires the attachment of electrodes on the skin, which is inconvenient.

	As an alternative, in this paper, it is proposed to use phonocardiogram (PCG) instead of ECG for the proximal timing reference for PTT measurements. Basically, PCG is produced due to the opening and closing of heart valves. There are two dominant types of sounds in each cardiac cycle, which are known as S1 and S2. 
\begin{figure}[!t]
\centering
\includegraphics[width=0.43\textwidth]{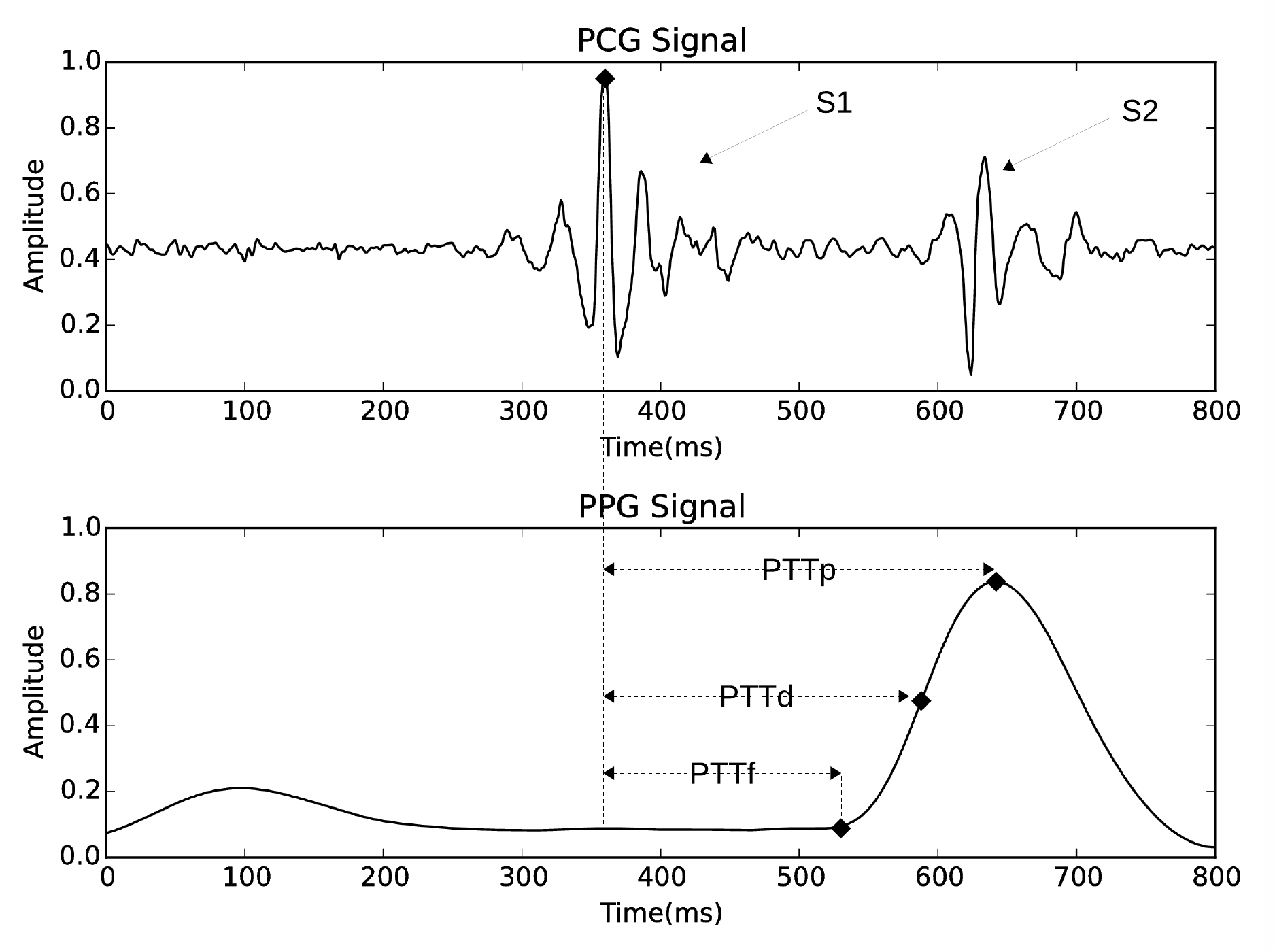}
\caption{{\footnotesize \textcolor[rgb]{0.00,0.00,0.00}{Variations of PTT values corresponding to different characteristic points on PPG.}
		}} \label{fig.signals}
\end{figure}
PCG S1-peak, instead of ECG R-peak, can be considered as the proximal timing reference for measuring PTT, as it physically represents the moment that blood leaves the heart and thus it is able to eliminate electromechanical delay of the ECG. Therefore, it can exclude PEP from the measurements to some extent. For the distal timing reference, a characteristic point on PPG, in the same cardiac cycle as the selected proximal point, is considered. This point could be either the point where PPG starts to rise in each cardiac cycle, or the point where the slope of PPG reaches its maximum, or the peak of PPG. PTT values corresponding to these points are called PTT$_{\textrm{f}}$, PTT$_{\textrm{d}}$ and PTT$_{\textrm{p}}$, respectively (see Fig. \ref{fig.signals}). Using PCG is very desirable as it requires a simple circuitry to be acquired (e.g. a microphone circuitry). 


\section{Data Collection Methodology} \label{section.datacollection}
\subsection{Hardware Setup} \label{section.equipment}
\begin{figure}[!t]
	\centering
	\includegraphics[width=0.45\textwidth]{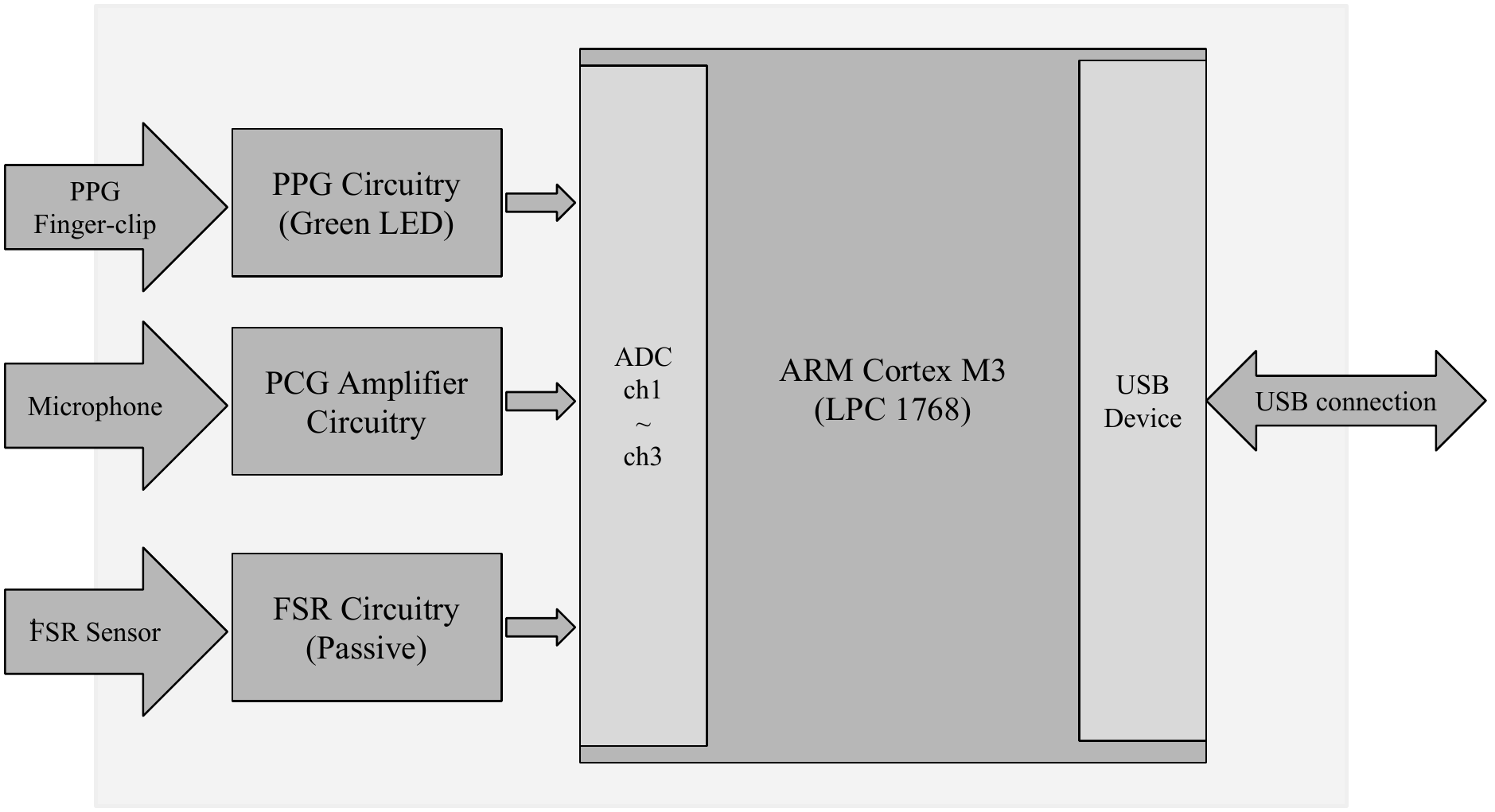}
	\caption{{\footnotesize \textcolor[rgb]{0.00,0.00,0.00}{Block diagram of the sampling hardware device.}}} \label{fig.hw}
\end{figure}

The hardware setup used in this work, consists of a data acquisition board that is designed in our team, a portable computer, and a commercial BP monitor. The board is responsible for acquiring the vital signals (i.e. PCG and PPG) as well as the signal from a Force Sensing Resistor (FSR). The signal recording is performed synchronously from three separate channels at the fixed sampling rate of 1KHz. The board is connected to the portable computer via Universal Serial Bus (USB) protocol and transfers data frames of 1-second length (see Fig. \ref{fig.hw}).
Moreover, a software is developed for the computer, which is responsible for saving the time series data from the channels and reference BP values.

In this setup, reflection PPG using 530 nm (green) wavelength was captured from the finger of the subject in the left hand. 
For PCG, a electret condenser microphone, which is connected to a diaphragm by means of a hollow tube was used. The diaphragm was placed on the subject’s chest and its location was fixed with a chest belt. Reference BP measurements during the calibration procedure, were taken from upper arm in the right hand with a digital automatic BP monitor (Model M3, OMRON, Japan).
FSR was placed under the inflatable cuff in order to measure the instantaneous applied pressure by the cuff. As the cuff inflates and deflates, using FSR is an effective and novel approach to distinguish the reference BP measurement time instances during the signal recording. 


\subsection{Data Acquisition Procedure} \label{section.experimentalsetup}
In total, 24 healthy subjects in the age range of 21-50 years were included in our experiments. 
For perturbing BP for the calibration procedure, all the subjects underwent a same
supervised physical exercise (running about 3 minutes at 8 km/h).
This physical exercise would cause their BP to rise in a discernible way. Immediately after the exercise, the subject would sit upright and the
data collection process starts. 
During data collection, the subjects were told not to speak and remain stable. For each subject, several reference BP measurements were taken.  Since each subject sits during the data collection after the exercise, the measured BP is generally decreasing through the measurements.

\begin{figure}[!t]
	\centering
	\includegraphics[width=0.43\textwidth]{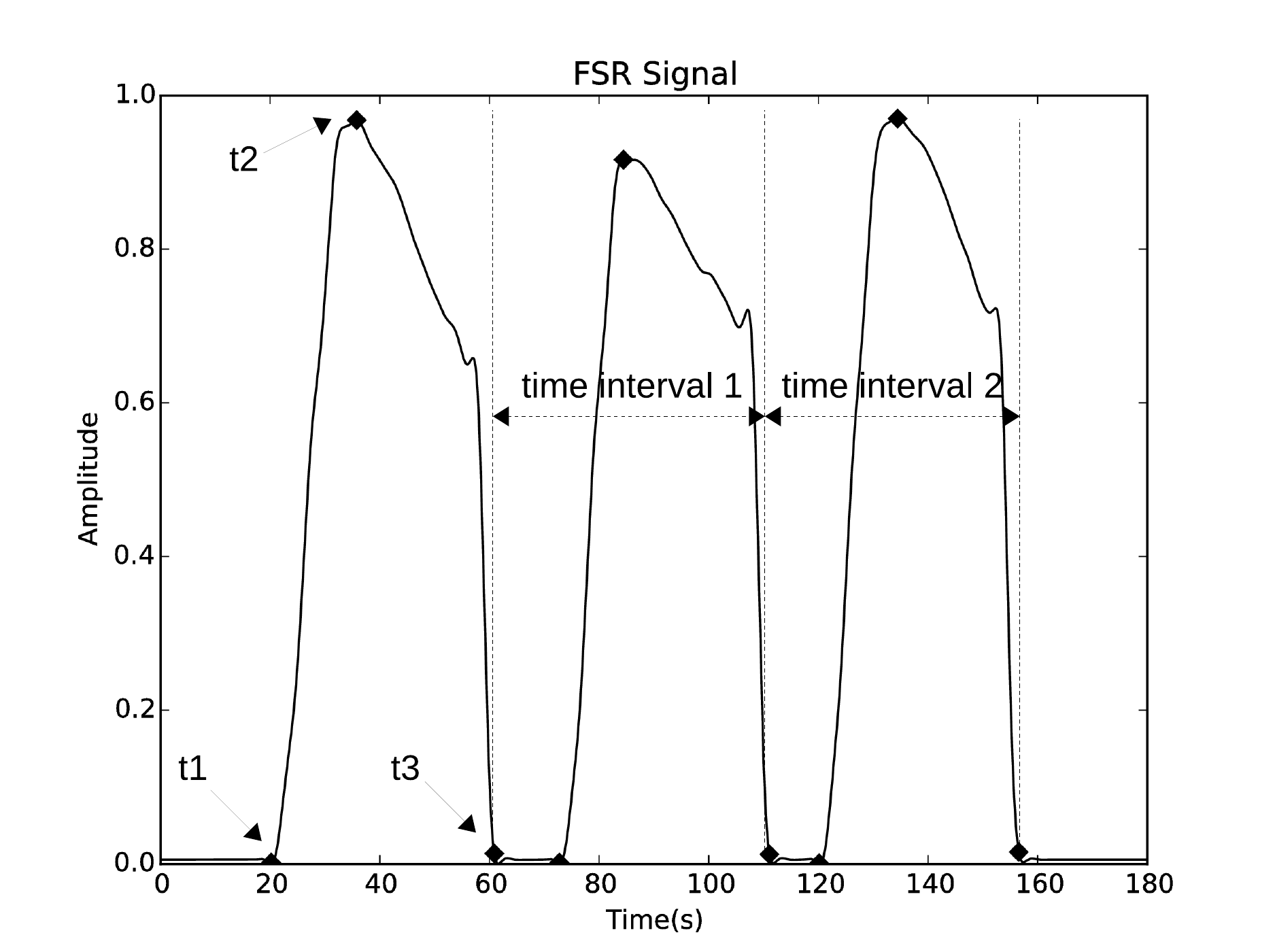}
	\caption{{\footnotesize The key moments on the FSR signal. Time intervals corresponding to $t_3$ moments are indicated.}} \label{fig.fsrdel}
\end{figure}

\section{Processing Pipeline} \label{section.processingpipeline}
\subsection{Preprocessing} \label{section.preprocess}
For the preprocessing purpose, at first, all four
signals are filtered using median filtering to remove the impulsive noise. Afterwards, a statistical normalization is
applied to each of the signals. At the end of this stage,
frequency selective filtering is done on normalized signals. The filter specifications for each of the acquired signals in the data collection phase are represented in Table \ref{table.filter}.
\begin{table}
\centering
\renewcommand{\arraystretch}{1.2}
\caption{Filter Specifications}\label{table.filter}
\begin{tabular}{|c|c|c|c|c|}

     \hline
  Signal  & Filter  & Low   & High   & Order  \\
     Name &  Type &  Cutoff  &  Cutoff  &    \\
     \hline
     FSR Signal  & IIR, Low-Pass & 0.3Hz  & -  & 3\\
     \hline
     PPG  & IIR, Band-Pass & 0.5    & 20  & 3\\
     \hline
     PCG  & IIR, Band-Pass & 20     & 240 & 3\\
     \hline

\end{tabular}
\end{table}

\subsection{PTT Measurement and Post-Processing} \label{section.PTT_PAT}
At first, all the signals are divided to time intervals corresponding to the reference BP measurements. In each measurement of BP using the automatic BP monitor, there are three key moments: 
the moment that cuff inflation starts ($t_1$), the moment that cuff deflation starts ($t_2$), the moment that cuff deflation ends and measured SBP and DBP values are read by the automatic BP monitor ($t_3$). These moments can be detected with the aid of the FSR signal, as it indicates instantaneous applied pressure by the cuff. The middle of consecutive $t_3$ values are used to divide the two main signals (i.e. PCG and PPG) to the desired time intervals (see Fig. \ref{fig.fsrdel}).
 
PTT values corresponding to each of reference BP measurements are calculated from corresponding time intervals.
An averaging is done on the extracted PTT values in each of the time intervals to provide one PTT value for each of the reference BP measurements in each time interval. Here, a weighted averaging is used in a way that puts more emphasis on PTT values closer to $t_3$ moments, since these moments indicate time instances when reference measurements are read.


\begin{table*}
  \centering

    \renewcommand{\arraystretch}{1.2}

  \caption{\textcolor[rgb]{0.00,0.00,0.00}{The SBP and DBP Estimation Results}}\label{table.results}
  \begin{tabular}{c|c|c|c|c|c|c|}
  \cline{2-7}
  & \multicolumn{3}{c|}{SBP (mmHg)} & \multicolumn{3}{c|}{DBP (mmHg)}\\
  \cline{2-7}
  \cline{2-7}
  &MAE&STD&r&MAE&STD&r\\

  \hline







\multicolumn{1}{|c|}{$PTT_f$}&10.33&14.94&0.71   	&4.40&5.67&0.78		\\

  \hline

  \multicolumn{1}{|c|}{$PTT_d$}&7.37&11.09&0.84   	&3.68&4.68&0.85		\\

\hline

  \multicolumn{1}{|c|}{$PTT_p$}&7.47&11.08&0.84   	&3.56&4.53&0.86	\\

  \hline
  \end{tabular}
\end{table*}

\section{Results and Discussions} \label{section.results}
For testing the models for each subject, leave-one-out separation
of the dataset to train and test sets is employed. It is ensured that there is no overlap between the train and test data.
Table \ref{table.results} presents the performance of the proposed method. Criteria for performance evaluation are mean absolute error (MAE), standard deviation (STD) and correlation coefficient (r) of the estimation.

\textcolor[rgb]{0.00,0.00,0.00}{According to Table \ref{table.results}, the estimation of the model when PTT$_{\textrm{f}}$ is used for PTT measurements is not as accurate as estimation using PTT$_{\textrm{d}}$ or PTT$_{\textrm{p}}$, where these two performs approximately the same for both SBP and DBP estimation. In this paper, PTT$_{\textrm{p}}$ is used for PTT measurements for the sake of the evaluation of results.}

\begin{table*}
	
	\centering
	\caption{Comparison with Other Works}
	\label{table.otherwork}
	\scalebox{0.9}{
		\begin{tabular}{|c|c|c|c|c|c|c|c|c|c|c|}
			\hline
			Work & Number of & \multicolumn{3}{c|}{Estimation Methodology} &
			\multicolumn{3}{c|}{SBP (mmHg)} & \multicolumn{3}{c|}{DBP (mmHg)}\\
			\cline{3-11}
			&Subjects&Signals&Index&Calibration Interventions&MAE&STD&r&MAE&STD&r\\
			\hline
			This Work &24&PCG \& PPG&PTT&Running Exercise&7.47&11.08&0.84&3.56&4.53&0.86\\
			\Xhline{2\arrayrulewidth}
			\cite{wong2009evaluation}&14&ECG \& PPG&PAT&Treadmill Exercise&-&5.3&0.87&-&2.9&0.30\\
			\hline
			\cite{gesche2012continuous}&63&ECG \& PPG&PAT&Cycle Ergometer Exercise&-&10.10&0.83&-&-&-\\
			\hline
			\cite{kim2015ballistocardiogram}&15&BCG \& PPG&PTT&Deep Breathing + Sustained Handgrip&&8.58 \footnotemark[1]&0.70&&5.81 \footnotemark[1]&0.66\\
			\hline
		\end{tabular}
	}
	
	\begin{flushleft}
		\hspace{0.15cm} \footnotemark[1] {\scriptsize Approximated from the corresponding Bland-Altman plot.\\}
	\end{flushleft}

\end{table*}




\begin{figure}
	\centering
	\begin{subfigure}[t]{0.43\textwidth}
		\centering
		\includegraphics[width = 1\textwidth]{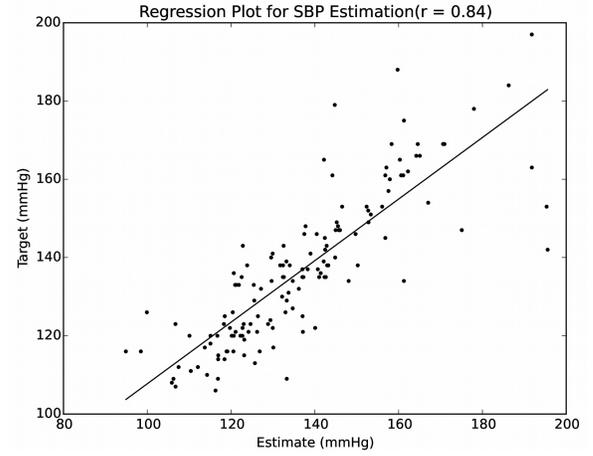}
		\caption{}
		\label{fig.reg_linear_sbp}
	\end{subfigure}
	\begin{subfigure}[t]{0.43\textwidth}
		\centering
		\includegraphics [width = 1\textwidth]{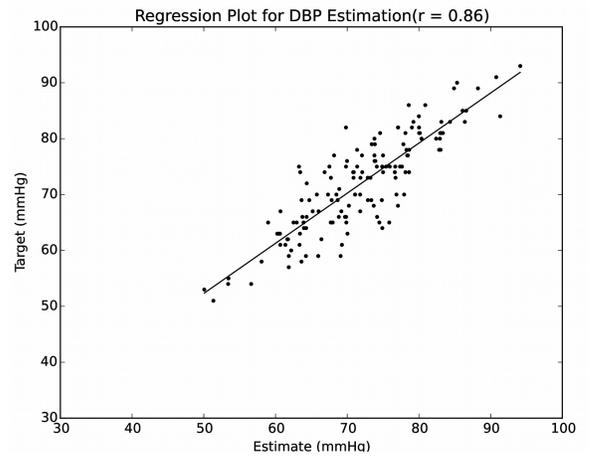}
		\caption{}
		\label{fig.reg_linear_dbp}
	\end{subfigure}
	\caption{\footnotesize{\textcolor[rgb]{0.00,0.00,0.00}{Regression Plots for  (a) SBP Targtes and (b) DBP Targets.}}}\label{fig.reg}
\end{figure}

\begin{figure}
	\centering
	\begin{subfigure}[t]{0.43\textwidth}
		\centering
		\includegraphics[width = 1\textwidth]{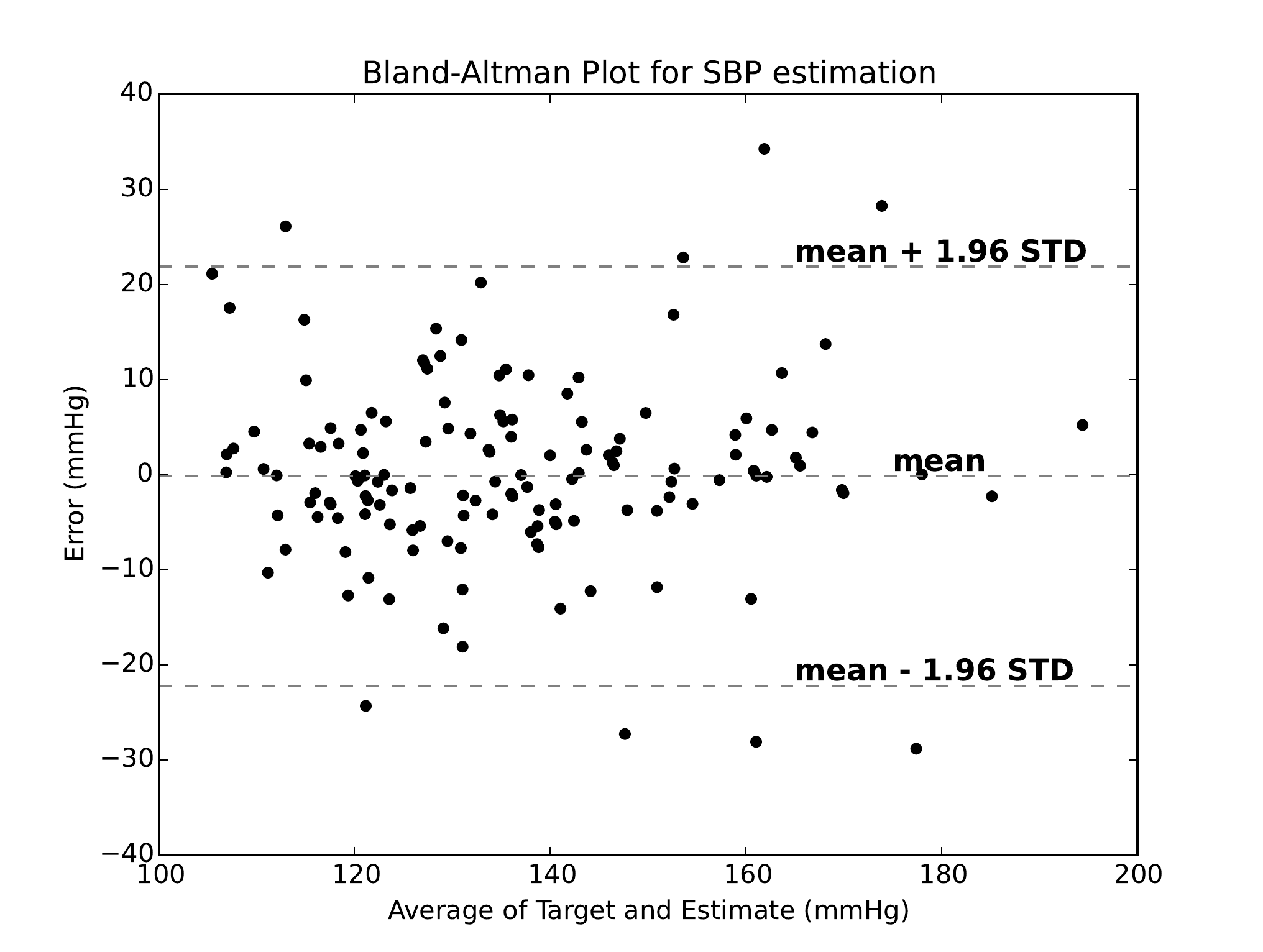}
		\caption{}
		\label{fig.bland_linear_sbp}
	\end{subfigure}
	\begin{subfigure}[t]{0.43\textwidth}
		\centering
		\includegraphics [width = 1\textwidth]{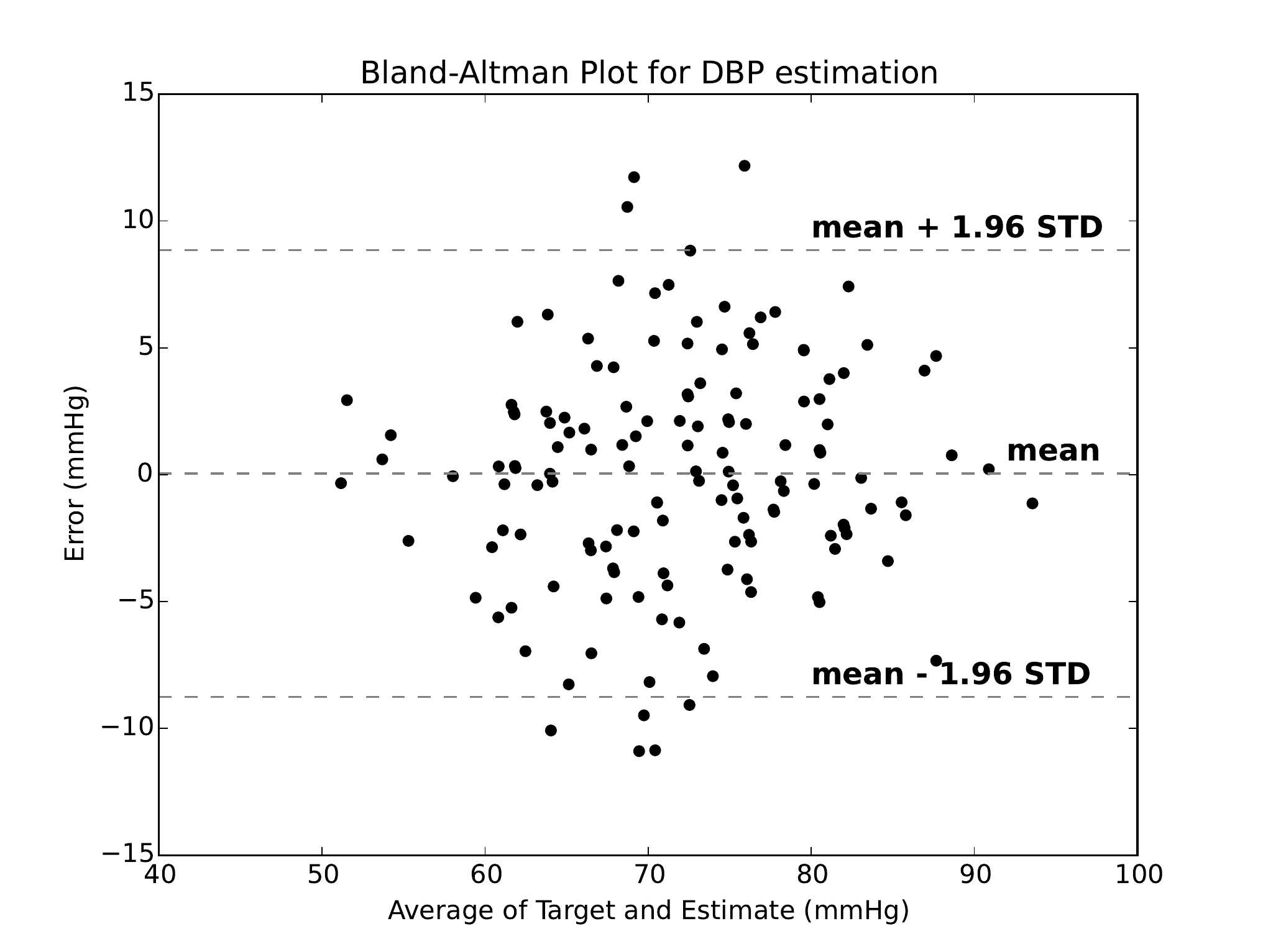}
		\caption{}
		\label{fig.bland_linear_dbp}
	\end{subfigure}
	\caption{\footnotesize{\textcolor[rgb]{0.00,0.00,0.00}{Bland-Altman Plots for  (a) SBP Targtes and (b) DBP Targets.}}}\label{fig.bland}
\end{figure}

The estimation versus target regression plots are presented in Fig. \ref{fig.reg}. The r value, for SBP and DBP estimations, calculated for the data of all 24 subjects, are 0.84 and 0.86, respectively. The achieved results prove that a strong correlation exists between estimation and target values for both SBP and DBP.

\textcolor[rgb]{0.00,0.00,0.00}{Fig. \ref{fig.bland} demonstrates the Bland-Altman plots for estimations. Based on Fig. \ref{fig.bland}, a large range of target values can be estimated with low errors (ranges for SBP and DBP targets are [106 mmHg , 197 mmHg] and [51 mmHg , 93 mmHg], respectively). Also, the limits of agreement (mean $\pm$ 1.96$\times$STD) are indicated, i.e., 0.16 $\pm$ 1.96$\times$11.08 mmHg for SBP and 0.04 $\pm$ 1.96$\times$4.53 mmHg for DBP.}

\subsection{Comparison with Other Works} \label{section.comparison}
Table \ref{table.otherwork} presents a comprehensive comparison between the performance of the proposed method in this paper (using PTT index) and a number of other works in literature, which use PAT or PTT to estimate BP. 
According to this table, in comparison with the methods which use alternatives other than PCG (e.g. ballistocardiogram (BCG) \cite{kim2015ballistocardiogram}) to measure PTT, the results for this work are more promising (r = 0.84 versus r = 0.70 for SBP estimation). The r value obtained in this work for the estimation of SBP, is close to the r value obtained in other works using the PAT index for the estimation of SBP (e.g., r = 0.87 in \cite{wong2009evaluation} and r = 0.83 in \cite{gesche2012continuous}). What is worthy to be mentioned is the fact that DBP estimations in this paper, have encouraging accuracies compared to other studies (by comparison of r values). This fact is more pronounced when we consider the fact that  in many studies in literature, correlation coefficients associated with DBP estimations are significantly less than those of in SBP estimations (see \cite{wong2009evaluation} and \cite{hennig2013continuous} for example). More importantly, the proof of the possibility of using PCG for BP estimation with comparable performance is one of the achievements of this paper.

\section{Conclusion} \label{section.conclusion}
In the present study, the problem of continuous and cuff-less BP estimation based on PTT was addressed. 
To achieve this goal, the proposed method in this paper, establishes a cuff-less BP estimation processing pipeline using vital signals such as PCG and PPG as inputs. This study showed that PCG signal can be employed for the purpose of BP estimation. This is desirable since PCG requires minimal low-cost equipment to be recorded and even commercial portable devices such as smartphones can be easily adapted for this purpose.


\end{document}